\documentclass[sigconf]{acmart}

\usepackage{mathtools}
\usepackage{subcaption}
\usepackage{caption}

\usepackage{algorithm}
\usepackage[noend]{algpseudocode}

\AtBeginDocument{%
  \providecommand\BibTeX{{%
    \normalfont B\kern-0.5em{\scshape i\kern-0.25em b}\kern-0.8em\TeX}}}

\setcopyright{acmcopyright}
\copyrightyear{2023}
\acmYear{2023}
\setcopyright{acmlicensed}\acmConference[WWW '23 Companion]{Companion Proceedings of the ACM Web Conference 2023}{April 30-May 4, 2023}{Austin, TX, USA}
\acmBooktitle{Companion Proceedings of the ACM Web Conference 2023 (WWW '23 Companion), April 30-May 4, 2023, Austin, TX, USA}
\acmPrice{15.00}
\acmDOI{10.1145/3543873.3584621}
\acmISBN{978-1-4503-9419-2/23/04}

\acmSubmissionID{21}

\begin{document}

\title{Explicit and Implicit Semantic Ranking Framework}


\author{Xiaofeng Zhu}
\email{xiaofzhu@microsoft.com}
\affiliation{%
  \institution{Microsoft}
  \city{Redmond}
  \state{WA}
  \country{USA}
  \postcode{98052}
}

\author{Thomas Lin}
\email{tlin@microsoft.com}
\affiliation{%
  \institution{Microsoft}
  \city{Redmond}
  \state{WA}
  \country{USA}
  \postcode{98052}
}

\author{Vishal Anand}
\email{visanand@microsoft.com}
\affiliation{%
  \institution{Microsoft}
  \city{Redmond}
  \state{WA}
  \country{USA}
  \postcode{98052}
}

\author{Matthew Calderwood}
\email{matthew.calderwood@nuance.com}
\affiliation{%
  \institution{Nuance Communications}
  \city{Burlington}
  \state{MA}
  \country{USA}
  \postcode{01803}
}

\author{Eric Clausen-Brown}
\email{erclause@microsoft.com}
\affiliation{%
  \institution{Microsoft}
  \city{Redmond}
  \state{WA}
  \country{USA}
  \postcode{98052}
}

\author{Gord Lueck}
\email{gordonl@microsoft.com}
\affiliation{%
  \institution{Microsoft}
  \city{Redmond}
  \state{WA}
  \country{USA}
  \postcode{98052}
}

\author{Wen-wai Yim}
\email{yimwenwai@microsoft.com}
\affiliation{%
  \institution{Microsoft}
  \city{Redmond}
  \state{WA}
  \country{USA}
  \postcode{98052}
}

\author{Cheng Wu}
\email{wucheng@microsoft.com}
\affiliation{%
  \institution{Microsoft}
  \city{Redmond}
  \state{WA}
  \country{USA}
  \postcode{98052}
}

\renewcommand{\shortauthors}{Zhu, et al.}

\begin{abstract}

The core challenge in numerous real-world applications is to match an inquiry to the best document from a mutable and finite set of candidates. Existing industry solutions, especially latency-constrained services, often rely on similarity algorithms that sacrifice quality for speed. In this paper we introduce a generic semantic learning-to-rank framework, Self-training Semantic Cross-attention Ranking $(sRank)$. This transformer-based framework uses linear pairwise loss with mutable training batch sizes and achieves quality gains and high efficiency, and has been applied effectively to show gains on two industry tasks at Microsoft over real-world large-scale data sets: Smart Reply (SR) and Ambient Clinical Intelligence (ACI). In Smart Reply, $sRank$ assists live customers with technical support by selecting the best reply from predefined solutions based on consumer and support agent messages. It achieves 11.7\% gain in offline top-one accuracy on the SR task over the previous system, and has enabled 38.7\% time reduction in composing messages in telemetry recorded since its general release in January 2021. In the ACI task, $sRank$ selects relevant historical physician templates that serve as guidance for a text summarization model to generate higher quality medical notes. It achieves 35.5\% top-one accuracy gain, along with 46\% relative ROUGE-L gain in generated medical notes.

\end{abstract}



\maketitle

\section{Introduction}

Learning-to-rank frameworks are versatile and extensible, especially in production environments where classification fails to scale or reaches performance limitations. In explicit information retrieval systems such as search engines, the design specifics of these frameworks are important - e.g., training and inference setup, data, and learning-to-rank model loss functions. In this paper we present a learning-to-rank framework $sRank$ developed for and tested on two industry tasks which both contain retrieval components with one binary relevance: Smart Reply (SR) and Ambient Clinical Intelligence (ACI). $sRank$ is tailored toward optimization and generalization to meet production requirements for these tasks.

With the popularity of deep neural ranking models in learning-to-rank \cite{huang2013learning,shen2014learning}, contextual featurization using Transformer models such as BERT \citep{devlin-etal-2019-bert} and Big Bird \cite{zaheer2020big} largely eliminate human efforts and achieve high performance in ColBERT \cite{khattab2020colbert}, PARADE \cite{li2020parade}, MVA \cite{zhou2021multi}, and more \cite{karpukhin2020dense,macavaney2020efficient,han2020learning}. To reduce inference cost for ranking, multi-stage retrieval systems can be set up where the top $k$ documents are first identified using ranking functions like BM25 \cite{robertson2009probabilistic}, and then neural models re-rank those $k$ documents. It is also common for production retrieval systems to use CPU for inference with relatively lightweight and less effective models, such as similarity of embedding spaces \cite{karpukhin2020dense}. However, these studies cannot fully address retrieval challenges for our SR and ACI tasks.

Smart Reply is a system for Microsoft's customer technical support chat that efficiently suggests reply messages for support agents serving multiple products. Its goal is to improve agent productivity, improve customer satisfaction, and reduce operation costs. Prefabricated/canned reply messages are created and reviewed by agent support specialists. Smart Reply then monitors conversations between customers and support agents, and suggests the top one canned reply message that agents can quickly use when the conversation context relates to a prepared reply. Product requirements for SR include that it must present suggestions faster than the agent's normal response and search time, it must suggest replies with low error tolerance, and it must provide smart replies only when needed to avoid overwhelming agents.

In the Ambient Clinical Intelligence task, visits between patients and physicians are recorded, transcribed by ASR systems, and then text generation models use this to automatically generate the needed medical note documentation for the encounter \cite{Enarvi2020GeneratingMR, Knoll2022UserDrivenRO, MedicalSum2022}. Generating medical documentation for physicians allows physicians to provide more attentive care to their patients, instead of being distracted by note taking and documentation during the encounter. It reduces physician burnout by saving physicians from many hours of documentation work. For medical documentation, physicians often re-use $templates$ that they have prepared beforehand for various encounter types. If a ranking model can select which one of the physician's templates is most appropriate for each new encounter transcript, then this information can be used to guide more accurate medical note generation. Product requirements for ranking in ACI include that it must select the correct template from a set of existing templates, and it must be computationally efficient both for training and inference.

Our $sRank$ framework addresses the challenges for SR and ACI. It effectively and efficiently applies a dual-encoder-style cross-attention architecture to learning-to-rank in both training and real-time prediction utilizing document embedding caches and self-training. We show a training technique that enables training over candidate sets of various sizes, and present an efficient method for making pairwise cross entropy linear for our applications. We also explain how sRank is able to return no result when there are no correct matches. The primary contributions of this paper are as follows:
\begin{itemize}
    \item We present $sRank$, an efficient self-training cross-attention learning-to-rank model that can be used for real-time applications, various loss functions, and is scalable to mutable batch sizes. Although we focus on binary relevance applications in this paper, it can also be applied to multi-level ranking or generic contrastive learning, such as Siamese networks \cite{roy2019siamese,thakur2021augmented} or triplet loss \cite{hoffer2015deep}. 
    \item We demonstrate that pairwise cross-entropy for one binary relevance is $O(n)$ time complexity using tensor calculation, while general RankNet \cite{burges2005learning} loss is $O(n^2)$ with a reduction to batch-level $O(n^2)$ in an open source learning-to-rank framework TFR-BERT \cite{han2020learning}. We reduce inference complexity by caching document embeddings thus self-train and update the embeddings during training.
    \item We present our ranking components optimized for the real world SR and ACI industry applications. We show 11.7\% to 35.5\% gain in top-one accuracy, as well as corresponding downstream application gains. These components can also be easily extended to additional industry applications. 
\end{itemize}
    
We launched an English release of Smart Reply to global regions in January 2021. We report relative improvement metrics over the previous system from 4-months of A/B testing conducted in 2020, and report online metrics from the recent 3-month period in 2022. For ACI, we began ranking research in July 2021 and the work is ongoing. For reasons including to protect customer privacy, this paper reports on relative metric improvements for both applications.

\begin{figure*}[pt]
\includegraphics[scale=0.55]{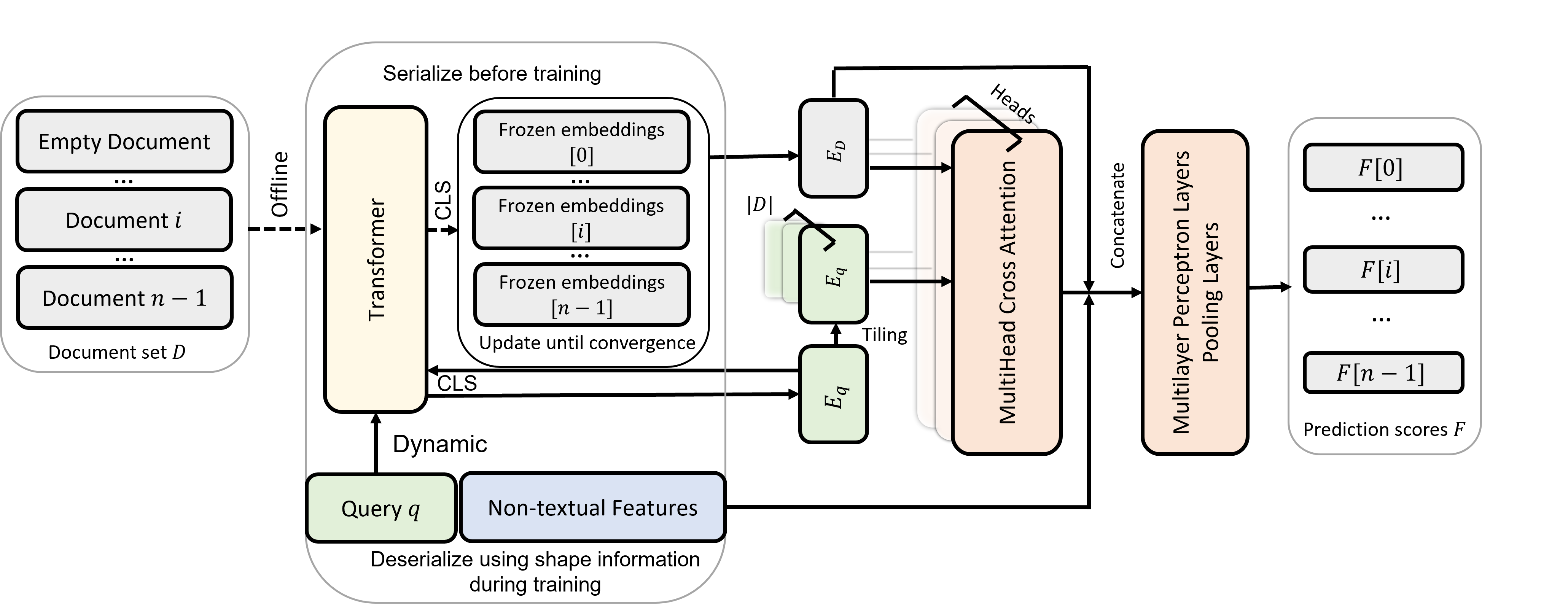}
\caption{The sRank model takes the serialized record of a query and the frozen embeddings of all its candidate documents before training, deserializes for Multi-Head Cross Attention, and generates the prediction scores $F$ for the candidate documents.}
\label{sRank model}
\Description[The sRank model]{The sRank model takes the serialized inputs associated with one query, deserializes them, and generate a list of prediction scores}
\end{figure*}

\section{Background and Related Work}
\label{Background and Related Work}

\subsection{Classical Learning-to-rank}

Learning-to-rank over classical and general retrieval systems with multi-level relevance (e.g., 0-5 with 0 being irrelevant and 5 being most relevant) often favors listwise loss functions over pairwise loss functions \cite{cao2007learning,xia2008listwise,wang2018lambdaloss}. Listwise loss functions are also chosen over pairwise loss functions for efficiency reasons, especially in neural networks, due to their $O(n)$ calculation instead of $O(n^2)$. However, for the retrieval components of our applications there is only one correct document per set of candidate documents. In Section \ref{Optimized pairwise Loss for Binary relevance} we show how tensor-based pairwise loss calculation can be optimized to $O(n)$ for our use cases.

Let $f(q, D^q)$ be a ranking function for ranking query $q$ and its associated candidate document set $D^q$ (we omit $q$ for simplicity in later sections). As there is only one correct answer for each set of candidate documents, $D^q$ = $d^{+} U D^{-}$, where $d^{+}$ is the document with label relevance of $1$, and $D^{-}$ indicates the rest of the candidate document with label relevance of $0$. NDCG differences in the gradients of LambdaRank and LambdaMart \cite{burges2010ranknet} or the loss functions of related work such as \cite{zhu2020listwise,NIPS2009_b3967a0e,taylor2008softrank,wang2018lambdaloss} become equivalent. We showcase representative RankNet and MLE-based loss functions that maximize the log likelihood of $P(d^{+})$ in Equations \ref{RankNet} and \ref{MLE} respectively.

\begin{equation} \label{RankNet}
P_t(d^+) = \frac{1}{|\{D^-\}|}\sum_{d^- \in \{D^-\}} \frac{1}{1 + e^{-(f(q, d^+) - f(q, d^-))}}.
\end{equation}

\begin{equation} \label{MLE}
P_t(d^+) = \frac{1}{1 + \sum_{d^- \in \{D^-\}} e^{-(f(q, d^+) - f(q, d^-))}}.
\end{equation}

The RankNet loss in Equation \ref{RankNet} is less likely to suffer gradient vanishing, and it can be implemented in $O(n)$ instead of batch-level $O (n^2)$ \cite{han2020learning} or strict $O(n^2)$ \cite{karpukhin2020dense}. In addition, current re-ranking models generally require truncating or padding the candidate set size to obtain a universal batch size for training. Section \ref{Optimized pairwise Loss for Binary relevance} shows how we can train and calculate loss effectively for complete sets of documents with varying sizes using mutable batch sizes.

\subsection{Transformer-based Re-ranking}

Prior studies such as DPR \cite{karpukhin2020dense} and PreTT \cite{macavaney2020efficient} define in-batch positive examples explicitly, while they draw in-batch negative examples from a larger pool in the training set, i.e., same negative examples are duplicated in different batches. In addition, training batches all have the same size. Unlike those studies, sRank has both an explicit positive example and explicit negative examples in a given batch, resulting in variably sized batches. sRank thus requires fewer training resources while obtaining a better pairwise training objective. We propose dual-encoder fashion cross-attention sRank that can be executed efficiently in real-time in Section \ref{sRank}.

\subsection{Application-Specific Requirements}

Support agent replies in Smart Reply are either generic and relevant for all products, or apply only for specific products. Clustering-based \cite{45189} and classification-based solutions do not satisfy product quality and scalability requirements for this scenario where answer sets may change or grow in this manner.

For the ACI template ranking task, physicians each have their own sets of historical templates, and do not choose templates from other physicians. Physicians can also have different numbers of templates, and the templates can be of varying sizes. Traditional re-ranking approaches are not ideal for this scenario because training batch size is fixed in neural models. A common solution is to truncate or pad candidate documents to a universal batch size in training. We show a generic way of training neural systems on batches of specific sets of candidates of mutable sizes. 

Lastly, relevance labels are binary for both SR and ACI, meaning that at most one reply or template can be presented. In this setting, pairwise sigmoid cross entropy is more suitable than listwise loss functions.

\section{Methodology}
\label{Methodology}
\subsection{Semantic Cross Attention Ranking}
\label{sRank}

The $query-key-value$ architecture of Multi-Head Cross Attention represented as $(E_q, E_D, E_D)$ in Figure \ref{sRank model} matches neatly with the expectations of text-based learning-to-rank studies by taking query embeddings/features $E_q$ as $query$ and document embeddings $E_D$ as $key$ and $value$. We cache the embeddings of candidate documents \cite{gao2020modularized} in the online system to reduce the inference latency from dual encoder with cross-attention. To ensure the inference approach works smoothly, we freeze the embeddings during training.

Instead of costlier methods such as training the embeddings on both queries and candidate documents or pre-training transformers using customized data, we utilize the learning-to-rank training process to update the weights in the transformer to generate more informative embeddings for the candidate documents. The embeddings of candidate documents are updated after several training epochs (e.g., 10) until the model converges. We apply multi-head cross attention to the frozen embeddings of candidate documents $E_D$ and the dynamic embeddings of each query $E_q$, tiled to $|D|$. 

The batch in sRank contains a serialized record of one question, and embeddings of its candidate documents and feature shapes using Parquet, enabling processing and training of candidate sets with mutable sizes. The negative examples come only from each Microsoft product's reply set in the SR task and each individual physician's template set in the ACI task. Negative examples are not duplicated, in contrast to DPR \cite{karpukhin2020dense}. Questions are not duplicated, in contrast to TFR-BERT \cite{han2020learning}. Therefore, our model saves training data and eliminates unnecessary computation of passing data to Transformer. We apply ONNX \cite{onnxruntime} quantization to inference.

\begin{algorithm}
\caption{Linear pairwise loss for one correct document}\label{Linear pairwise loss for one correct document}
\begin{algorithmic}[1]
\State \textbf{Input}: labels $Y=(Y_i)_{i=1}^n$
\State \textbf{Input}: prediction scores $F=f(q, (d_i)_{i=1}^n)$
\State $P\_DIFF \gets F - F^T$
\State $L\_DIFF \gets P\_DIFF \cdot Y$
\State $S \gets exp(L\_DIFF)$
\State $loss \gets -\frac{1}{n - 1} \sum ((1 - Y) \odot ln \frac{1}{1 + S})$
\State $loss \gets \frac{-ln 2 +\sum ln (1 + S)}{n - 1} $ \algorithmiccomment{only one correct document in $Y$}
\State \textbf{Return} $loss$
\end{algorithmic}
\end{algorithm}
\vspace{-6mm}

\subsection{Optimized Loss for Binary Relevance}
\label{Optimized pairwise Loss for Binary relevance}

Algorithm \ref{Linear pairwise loss for one correct document} shows the $O(n)$ tensor-based pairwise loss calculation optimized for our use cases. Matrix $P\_DIFF$ of size $n \times n$ contains the pairwise prediction score differences and vector $L\_DIFF$ of size $n \times 1$ contains the linear score differences between all of the candidate documents and the correct document. We can split candidate documents into multiple batches with the correct document in each batch when $n$ is too large for GPU memory. 

\section{Use cases and Experiments}
\label{Real-world use cases and Experiments}

This section describes our experiments and measurements for ranking in SR and ACI. Our primary metric in offline evaluation is top-one accuracy because we suggest at most one reply message for SR and at most one physician template for ACI. The proposed loss is 2-7\% better than the MLE loss from \cite{zhu2020listwise} on the two tasks.

\subsection{Smart Reply for Customer Support}

The Smart Reply task is to take the most recent support agent message and the most recent customer message in a customer support conversation, and choose the best reply from a set of canned reply templates. An example Smart Reply could be "This \underline{link} has step-by-step instructions for how to activate Microsoft 365." An efficient CPU-based classifier is applied at runtime to classify which specific product an incoming support message interaction is for, and then our learning-to-rank model is used to select the best reply from the canned replies for that product. Smart Reply is able to support customer support interactions across 22 Microsoft products such as Office 365, Teams, Surface, and Remote Assistance.



Table \ref{SR datasets} describes statistics for the Smart Reply task. Training and test data were based on an 80\%:20\% split ratio. Customized tokenization was first applied to message pairs and canned replies, and then DistilBERT \cite{sanh2019distilbert} was used to vectorize the queries and documents for ranking. One difference from traditional retrieval systems which always retrieve the top-k documents is that we do not want to overwhelm support agents with replies when there are no good canned replies for a customer message. To achieve this, we added a "Silent" class in the product classifier and an "Empty" canned reply in each candidate reply set. We used data augmentation to generate synthetic conversations. For instance, appending a message pair that returns the "Silent" class or "Empty" reply to non-empty questions enriched non-empty triplets, and only using agent or customer messages further enlarged the data size. 




\vspace{-2mm}
\begin{table}[ht]
\centering
\scalebox{0.8}{
\begin{tabular}{l|r}
\hline
Cleaned customer-agent message pairs  & 1.3 Million \\ \hline
Maximum input tokens  & 512 \\ \hline
Canned reply templates & 200 \\ \hline
Supported Microsoft products  & 22  \\ \hline
Canned reply templates per product & 3-26  \\ \hline
Data set size with augmentation  & 10 Million \\ \hline
\end{tabular}
}
\caption{SR data statistics}
\label{SR datasets}
\vspace{-11mm}
\end{table}

\vspace{-2mm}
\begin{table}[ht]
\centering
\scalebox{0.8}{
\begin{tabular}{l|r}
\hline
Top-one accuracy gains  & 11.7 \\ \hline
Click-through rate (CTR) uplift & 42.5 \\ \hline
Agent satisfaction improvement  & 13.4 \\ \hline
Time reduction for composing agent messages & 38.7\\ \hline
\end{tabular}
}
\caption{SR offline and online metric gains (\%)}
\label{SR metric gains}
\vspace{-9mm}
\end{table}




Table \ref{SR metric gains} shows the 11.7\% offline top-one accuracy gain of sRank compared to our previous DSSM-based \cite{huang2013learning} system that took transformer embeddings as inputs. We exposed Smart Reply to insider agents for initial feedback then to 50\% global agents during A/B testing. sRank also increased CTR on Smart Replies by an absolute 42.5\%. During A/B testing, Smart Reply with sRank led to 13.4\% increase in agent satisfaction compared to the group not using Smart Reply, and agents composed replies 38.7\% faster with Smart Reply.

\subsection{Template Ranking in ACI}

The second industry task we tackle is Template Ranking in Ambient Clinical Intelligence. Physicians can have sets of templates that they start from when composing medical documentation. An example of a template could be: "General Appearance: Height \_\_ inches. Weight \_\_ pounds. The patient is alert and oriented and in no distress.." If a template ranking system can automatically select the best template to use for an encounter, then this can be used to guide the automatic generation of a more accurate AI medical note.


Table \ref{ACI encounter data} describes statistics for the ACI task. For this task the query is the medical encounter transcript and the candidate documents are physician's templates plus an "Empty" template to represent if no template should be used. We utilized Big Bird RoBERTa to generate conversation and template embeddings. Once our ranking model selected a template for an encounter, the template and the encounter transcript were concatenated and passed to the note generation model to generate the medical note. We focused on guiding generation of the Physical Exam section of Orthopedics clinical notes, because this section often employs templates. The baseline system is a DPR ranking model which meets the inference time requirements for this task.



\vspace{-2mm}
\begin{table}[ht]
\centering
\scalebox{0.8}{
\begin{tabular}{l|r}
\hline
Medical encounters for ranker training  & 1 Million+ \\ \hline
Maximum input tokens  & 4096 \\ \hline
Total number of medical templates & 6118  \\ \hline
Medical templates per physician & 8-39  \\ \hline
\end{tabular}
}
\caption{ACI template modeling configuration}
\label{ACI encounter data}
\vspace{-9mm}
\end{table}

\vspace{-1mm}
\begin{table}[ht]
\centering
\scalebox{0.8}{
\begin{tabular}{l|r}
\hline
ROUGE-L of baseline relative to ROUGE-L with oracle templates & 46.0 \\ \hline
ROUGE-L of sRank relative to ROUGE-L with oracle templates & 92.0 \\ \hline
Top-one accuracy gain over DPR &35.5 \\ \hline
Top-one accuracy gain (<25\% new templates) &41.5 \\ \hline
Top-one accuracy gain (25-75\% new templates) &40.6 \\ \hline
Top-one accuracy gain (>75\% new templates) &20.7 \\ \hline
\end{tabular}
}
\caption{ACI sRank metric gains (\%)}
\label{sRank metric}
\vspace{-7mm}
\end{table}


When medical note generation is guided by the correct template that the physician would use (the "oracle" template), this significantly increases the quality of generated medical notes. The challenge is then how to predict the correct template at run time. Note generation with no template guidance achieved only 46\% of the ROUGE-L \cite{lin2004rouge} of using oracle guidance. The DPR model was unable to predict templates at sufficient accuracy, and using its templates led to 2\% ROUGE-L drop compared to using no templates. sRank predicted templates more accurately, and end-to-end ROUGE-L with sRank was 92\% of oracle ROUGE-L. This highlighted the ability of sRank to effectively guide the generation of higher quality medical notes. Table \ref{sRank metric} shows sRank metrics for ACI.

For ranking metric gains, sRank achieved 35.5\% higher top-one accuracy than our DPR model with 7.5\% less inference time. To evaluate robustness of sRank to template editing (the scenario where a physician further edits their templates after ranker training), we also verified that it achieved accuracy gains across subsets of physicians whose test encounters contain edited templates at various frequencies. 





\section{Conclusion}
\label{Conclusion}

This paper presents our cross-attention learning-to-rank sRank model that we developed for two real-world industry tasks at Microsoft. We describe a number of optimizations and improvements that enable sRank to perform better on our tasks than previous models. The high quality and speed of sRank may make it an attractive option for additional industry ranking tasks that require selecting top-one options from candidate sets.

%


\bibliographystyle{ACM-Reference-Format}
\bibliography{ref.bib}


\end{document}